\documentclass[lettersize,journal]{IEEEtran}
\usepackage{cite}
\usepackage{amsmath,amssymb,amsfonts,mathtools,bm,bbm,xfrac}
\usepackage{graphicx}
\usepackage{textcomp}
\usepackage{mathbbol}
\usepackage{pifont}
\usepackage{enumitem}
\setlist[itemize]{leftmargin=*}
\setlist[enumerate]{leftmargin=*}
\usepackage{soul}

\let\oldFootnote\footnote
\newcommand\nextToken\relax

\renewcommand\footnote[1]{%
    \oldFootnote{#1}\futurelet\nextToken\isFootnote}

\newcommand\isFootnote{%
    \ifx\footnote\nextToken\textsuperscript{,}\fi}
\PassOptionsToPackage{hyphens}{url}
\usepackage{hyperref}
\hypersetup{
    colorlinks=true,
    linkcolor=blue,
    filecolor=magenta,      
    urlcolor=cyan,
    pdfpagemode=FullScreen,
    citecolor=teal
    }
\usepackage{booktabs}
\usepackage[]{cleveref}
\usepackage[final]{changes}
\makeatletter
\AddToHook{cmd/added/before}{\def\Changes@AuthorColor{blue}}
\AddToHook{cmd/deleted/before}{\def\Changes@AuthorColor{red}}
\AddToHook{cmd/replaced/before}{\def\Changes@AuthorColor{cyan}}
\makeatother

\AtBeginEnvironment{appendices}{\crefalias{section}{appendix}}
\def\BibTeX{{\rm B\kern-.05em{\sc i\kern-.025em b}\kern-.08em
  T\kern-.1667em\lower.7ex\hbox{E}\kern-.125emX}}
  
\usepackage{amsthm}
\newtheorem{thm}{Theorem}

\newtheorem{defn}{Definition}

\newtheorem{remark}{Remark}

\crefname{lemma}{lemma}{lemmas}
\Crefname{lemma}{Lemma}{Lemmas}
\crefname{thm}{theorem}{theorems}
\Crefname{thm}{Theorem}{Theorems}
\crefname{remark}{remark}{remarks}
\Crefname{remark}{Remark}{Remarks}
\crefname{defn}{definition}{definitions}
\Crefname{defn}{Definition}{Definitions}
\crefname{cor}{corollary}{corollaries}
\Crefname{cor}{Corollary}{Corollaries}

\usepackage{algorithm}
\usepackage[]{algpseudocode}

\newcommand{\qtilde}[0]{\tilde{q}}

\newcommand{\XX}{\mbox{\tiny \it zz'}}

\newcommand{\X}{\mbox{\tiny \it X}}

\newcommand{\qz}{\mathbbm{q}(\bm{z})}
\newcommand{\tqz}{\tilde{\mathbbm{q}}(\bm{z})}
\newcommand{\Hza}{\mathcal{H}_0^{(a)}}
\newcommand{\Hoa}{\mathcal{H}_1^{(a)}}

\DeclareMathOperator*{\argmin}{arg\,min}

\usepackage{subfig}

\begin{document}
\title{\deleted{Preserving Smart Grid Integrity: A }Differential\added{ly} Priva\replaced{te}{cy} \replaced{Communication of Measurement Anomalies}{Framework for Secure Detection of False Data Injection Attacks} in the Smart Grid}

\author{
Nikhil~Ravi,~\IEEEmembership{Graduate Student Member,~IEEE},
Anna~Scaglione,~\IEEEmembership{Fellow,~IEEE},
% Julieta Giraldez,~\IEEEmembership{Member,~IEEE},
Sean~Peisert,~\IEEEmembership{Senior Member,~IEEE},
% Chuck Moran,
Parth~Pradhan,~\IEEEmembership{Senior Member,~IEEE}% <-this % stops a space
\thanks{N. Ravi and A. Scaglione are with the Department of Electrical and Computer Engineering, Cornell Tech, e-mail: nr337@cornell.edu.
S. Peisert is with Lawrence Berkeley National Laboratory.
P. Pradhan is with Kevala.
}
\thanks{
The Director, Cybersecurity, Energy Security, and Emergency Response, Cybersecurity for Energy Delivery Systems program, of the U.S. Department of Energy, under contract DE-AC02-05CH11231 supported this research.  Any opinions, findings, conclusions, or recommendations expressed in this material are those of the authors and do not necessarily reflect those of the sponsors of this work.}%
}% <-this % stops a space


\markboth{Journal of \LaTeX\ Class Files,~Vol.~14, No.~8, August~2015}%
{Shell \MakeLowercase{\textit{et al.}}: Bare Demo of IEEEtran.cls for IEEE Journals}



\maketitle

\begin{abstract}
  In this paper, we present a framework based on differential privacy (DP) for querying electric power measurements to detect system anomalies or bad data\deleted{ caused by false data injections (FDIs)}. Our DP approach conceals consumption and system matrix data, while simultaneously enabling an untrusted third party to test hypotheses of anomalies, such as \replaced{the presence of bad data}{an FDI attack}, by releasing a randomized sufficient statistic for hypothesis-testing. We consider a measurement model corrupted by Gaussian noise and a sparse noise vector representing the attack, and we observe that the optimal test statistic is a chi-square random variable. To detect possible attacks, we propose a novel DP chi-square noise mechanism that ensures the test does not reveal private information about power injections or the system matrix. The proposed framework provides a robust solution for detecting \replaced{bad data}{FDIs} while preserving the privacy of sensitive power system data.
\end{abstract}


% \fi
\IEEEpeerreviewmaketitle

\section{Introduction}
\begin{figure}[!htbp]
  \centering
  \includegraphics[width=0.4\textwidth]{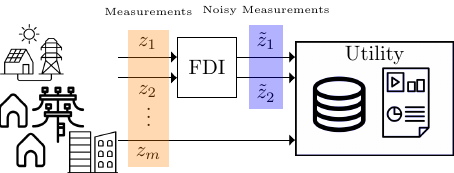}
  \caption{Illustration of bad data attacks.}
  \label{fig:fda_schematic}
\end{figure}

\IEEEPARstart{P}{ower} systems \replaced{are pivotal in delivering electricity to various sectors, including residential, commercial, and industrial. The effective management of these systems is heavily reliant on the accurate and timely acquisition of data for operational, control, and monitoring purposes~\cite{kalyani2011particle,wu2017data, dong2017short, zor2017state}. Such data is essential for numerous critical functions like load forecasting and security assessments. However, as power systems become increasingly digitized and interconnected, they are also more vulnerable to cyber threats. These include the malicious manipulation of measurement data, known as bad data (see \cref{fig:fda_schematic} for an illustration), which can lead to significant disruptions in operations, financial losses, and even endanger public safety~\cite{liu2011false}.  Consequently, the detection of such malicious activities emerges as a crucial area of focus, underpinning the need for robust cybersecurity measures within the fragmented smart grid.}{are critical infrastructures that supply electricity to households, businesses, and industries. Efficient management of power systems relies on accurate measurement data for monitoring and optimizing operational and control decisions. This data is also crucial for tasks such as short-term load forecasting~\mbox{\cite{wu2017data, dong2017short, zor2017state}} and security assessments~\mbox{\cite{kalyani2011particle}}. However, the increasing reliance on technology and interconnected systems has made power systems more vulnerable to cyber-attacks, particularly false data attacks (FDAs).}

\deleted{FDAs involve malicious manipulation of data or measurements (see \mbox{\cref{fig:fda_schematic}} for an illustration) to deceive power system control and operation, leading to erroneous decisions. These attacks may pose severe consequences, including blackouts, equipment damages, and financial losses, affecting the lives of millions of people~\mbox{\cite{liu2011false}}.}

\deleted{Motivations behind FDAs in power systems can vary, ranging from financial gain to political objectives and even cyber warfare. Attackers may seek to manipulate the energy market by creating shortages or price spikes or to inflict damage on power system infrastructure, thereby disrupting services and compromising public safety. Additionally, FDAs can be part of larger cyber warfare strategies, where the power system is exploited as either a target or a tool to achieve military or geopolitical objectives.}

\deleted{FDAs can occur at different stages of the power system, from generation and transmission to distribution. Attackers employ various techniques, such as injecting false data, replaying legitimate data, or executing man-in-the-middle attacks, to manipulate measurements, control signals, or communication networks used in power system operations. For instance, by injecting false measurements into power system sensors, attackers can mislead the system's state estimation process, which is crucial for determining the operating conditions.}

\deleted{These attacks can target specific power system components, such as generators, transformers, or protection relays, or can aim at disrupting the entire power system, resulting in cascading failures and widespread outages. Moreover, FDAs can be launched remotely via the Internet or by insiders with privileged access to power system equipment and networks.}

\subsection{\added{Motivation: Collective Defense}}

\replaced{T}{On the other hand, t}he management of power grids often suffers from fragmentation among multiple \replaced{Regional Transmission Organizations (RTOs)}{operators} \added{and utilities} and a division between distribution and transmission, despite the interconnected nature of the grid. Efficient prevention and detection of \replaced{cyber-physical attacks}{FDAs} necessitate collaboration and information sharing among different utilities. \added{The advent of Distributed Energy Resources (DERs) has further complicated the cybersecurity landscape of power systems. The decentralized nature and a mix of ownership between utilities and private stakeholders of DER-rich grids necessitates a collective approach to cyber-defense, emphasizing the importance of collaboration among various stakeholders. This \textit{collective defense} strategy is foundational to ensuring the secure and efficient operation of interconnected systems and devices.} \replaced{Furthermore, t}{T}his collaborative approach can significantly enhance early detection capabilities, improve understanding of attack methods, develop effective defense mechanisms, implement cost-effective solutions, and ensure regulatory compliance, ultimately bolstering the security and reliability of power systems. Anomalies in local data can serve as warning signs for issues that may have implications for neighboring systems~\added{\cite{saha2021secure}}, yet \added{achieving such a collaborative cybersecurity framework in} the fragmented \deleted{nature of} grid management \added{infrastructure} \replaced{is fraught with challenges, particularly in the realms of information sharing, technological solutions, collaborative partnerships, and coordinated incident response~\cite{liu2012cyber}.}{impedes the seamless exchange of data.}

\replaced{Information sharing is crucial for collective defense but is hindered by several factors, including concerns about data privacy related to customers, proprietary issues, and security~\cite{wh2024national}. Many grid operators are hesitant to share sensitive measurement data, such as Advanced Metering Infrastructure (AMI) and Phasor Measurement Units (PMU), further exacerbated by the fragmented landscape of grid devices and systems. This landscape now encompasses Home/Building Energy Management Systems, DER aggregators, and IoT devices, complicating the construction of a comprehensive and effective cyber-defense posture. Moreover, even sharing data with law enforcement agencies and regulators can encounter obstacles due to privacy considerations. This fragmentation and the myriad of concerns lead to critical cybersecurity information remaining siloed or entirely unshared, significantly compromising the grid's overall security and underscoring the importance of safeguarding against unintended disclosure of private data within the power system sector.}{Despite the potential benefits, sharing information encounters challenges due to concerns about data privacy related to customers, proprietary issues, and security. Furthermore, even sharing data with law enforcement agencies and regulators can face obstacles due to privacy considerations. Safeguarding against unintended disclosure of private data is therefore of significant importance in the power system sector.}

% \added{Technological limitations further obstruct the realization of an effective collective cyber-defense framework. Many utilities lack the necessary monitoring, control, and management capabilities to fully comprehend the grid's physical and cyber aspects. This gap in visibility is not limited to utility-owned devices and systems but extends to non-utility-owned entities, creating significant blind spots in the grid's cyber-defense. Moreover, the grid's security capabilities are often limited by the underutilization of advanced technologies, such as artificial intelligence and machine learning, which could otherwise enhance cybersecurity measures through proactive and automated approaches.}

\added{
% The absence of a unified platform for collaboration among utility leaders, similar to the Five-eyes intelligence alliances~\cite{pfluke2019history}, highlights significant challenges in cybersecurity intelligence sharing and incident response within the energy sector.
Without a dedicated platform for real-time information exchange and automated decision-making, utilities struggle to identify new threats, learn from incidents in other territories, and coordinate responses to cyber-attacks. This lack of structured incident response, coupled with insufficient knowledge of protecting operational technology systems, leaves the grid vulnerable to sophisticated cyber-attacks. Recognizing these challenges, the United States government emphasizes the importance of technology software equipped with a collective defense capability that enables rapid sharing of insights and detections with the Federal government, participants, and other trusted (by the US government) organizations, thereby enhancing grid resilience against sophisticated cyber-attacks~\cite{ceser2023considerations}. Moreover, the recent initiatives by the White House and the Department of Energy to fund research into innovative cyber-physical collective defense methodologies are timely and crucial. This funding aims to foster the development of defense strategies that are effective even in scenarios involving potentially untrustworthy third parties, thereby signaling a significant step towards bolstering the resilience and integrity of our energy infrastructure~\cite{wh2021national,doecollectivedefense}.}

\added{In this context, differential privacy (DP)~\cite{dwork2006our} mechanisms emerge as a promising solution to the challenges faced in implementing a collective defense strategy. By enabling the secure sharing of data among grid operators and with third-party entities, DP addresses key concerns around data privacy and proprietary information. DP mechanisms, by introducing controlled statistical ``noise'' to the data, protect sensitive information while maintaining its utility for analytical purposes, offering a balance between privacy and accuracy. This method surpasses traditional anonymization techniques by providing mathematical guarantees on the amount of information leakage, allowing for the optimization of queries relevant to the energy sector and the design of differentially private databases for analysis and research. Such practicality in applying DP to energy datasets fosters increased data sharing, enhancing stakeholder comfort and safeguarding privacy, trade secrets, and sensitive information. Before summarizing our contributions, next, we provide a brief review of the literature on anomaly detection, DP, and DP anomaly detection for smart grid applications.}

\subsection{\added{Literature Review}}\label{sec:lit_review}
\subsubsection{\added{Anomaly Detection in the Smart Grid}}
\added{Efforts to combat bad data attacks on power systems have led to the development of a diverse array of detection algorithms. Data-driven strategies, such as those employing machine learning techniques like distributed Support Vector Machines for stability-focused detection~\cite{esmalifalak2014detecting}, and real-time electricity theft detection~\cite{jindal2016decision}, leverage large datasets to identify anomalies indicative of false data injection (FDI) attacks. Anomaly identification techniques utilizing Multiclass SVMs have shown efficiency \cite{khan2021intelligent}, though their computational demand limits broader application. Artificial neural networks and their extensions into deep learning have gained popularity for their high detection accuracy in identifying FDI attacks~\cite{habibi2020false,liu2021adaptive,ferragut2017real,habibi2020detection,dehghani2020deep,zhang2020detecting}, yet they suffer from extensive training times. Attempts to mitigate these computational challenges have led to innovative solutions, such as the integration of artificial bee colony algorithms with differential evolution theory \cite{yang2017improved}. Despite the advantages of data-driven methods, their effectiveness is curtailed by the need for extensive, often centralized, datasets, leading to challenges in time, cost, and privacy. The reliance on large local datasets or detailed system information introduces substantial data transmission burdens, while the absence of effective privacy-preserving measures and the risks associated with centralized data processing highlight the need for a more collaborative detection approach.}

\subsubsection{\added{DP for Smart Grids}}
\deleted{Previous approaches, such as access control~\mbox{\cite{ruj2013decentralized,wen2021feddetect,chang2023practical}} and anonymization~\mbox{\cite{efthymiou2010smart}}, have been explored, but they have limitations. Access control methods often provide either unrestricted or no access at all, while anonymization techniques are vulnerable to reidentification attacks~\mbox{\cite{Narayanan2008-short}}. In the case of electric grid data, regulators have proposed policies, such as the ``15/15 Rule''~\mbox{\cite{15_15rule}}, for sharing electric consumer data in the public domain. However, these rules lack scientific rationale and fail to provide adequate privacy guarantees, as demonstrated in~\mbox{\cite{ravi2022differentially}}. 
DP mechanisms offer provable privacy and accuracy trade-offs, enabling the optimization of queries relevant to the energy sector and the design of differentially private databases for analysis and research. By introducing controlled statistical ``noise'' to the data, DP mechanisms protect sensitive information while maintaining data utility for analytical purposes. Unlike anonymization techniques, DP mechanisms provide mathematical guarantees on the amount of information leaked to data analysts or other parties. By allocating privacy budgets, DP mechanisms limit maximum information leakage over a set of queries, providing approximate statistical answers and analyses optimized for utility and acceptable privacy leakage. The practicality and availability of general DP mechanisms present an opportunity to analyze cybersecurity data for energy delivery systems while preserving privacy. Applying DP mechanisms to energy datasets can increase stakeholders' comfort with data used for various analytical and planning purposes, leading to increased data sharing while safeguarding privacy, trade secrets, and sensitive information.
In recent years, the application of DP has gained traction in the domain of smart grids, showcasing its versatility and effectiveness in addressing privacy concerns. Prior research has explored the integration of DP mechanisms in various aspects of smart grid data management. For instance, DP has been employed in the reporting of demand data, ensuring that individual consumption patterns remain confidential while providing aggregated information for grid optimization~\mbox{\cite{tran2022smart}}. Additionally, DP has found applications in clustering load profiles, where the goal is to group similar consumption patterns without compromising the privacy of individual users~\mbox{\cite{ravi2022differentially}}. Furthermore, the same paper also discusses how publishing load profiles can be tackled using DP techniques to enable data sharing for research purposes without revealing sensitive information about specific consumers.} 

\added{The landscape of cybersecurity within power systems has seen various strategies aimed at preserving data privacy and enhancing system resilience. Traditional approaches such as access control and anonymization have been instrumental in initial efforts to protect sensitive data. Access control methods, as discussed in~\cite{ruj2013decentralized, wen2021feddetect, chang2023practical}, attempt to regulate data access, often resulting in binary outcomes of either complete access or total restriction. Meanwhile, anonymization techniques, highlighted in~\cite{efthymiou2010smart}, seek to obscure personal identifiers within datasets. Despite their intentions, these techniques have been criticized for their susceptibility to reidentification attacks, a vulnerability exposed by~\cite{Narayanan2008-short}. Regulatory attempts to navigate the privacy challenges inherent in electric grid data management, such as the ``15/15 Rule''\cite{15_15rule}, aim to balance consumer privacy with the public's right to access data. However, the effectiveness of these policies has been questioned, with critiques pointing out a lack of scientific underpinning and insufficient privacy safeguards, as elaborated in~\cite{ravi2022differentially}.}

\added{In response to these challenges, DP has emerged as a robust alternative, particularly in the context of smart grids. DP's capacity to protect individual privacy while enabling aggregated data analysis offers a solution to the limitations of previous approaches. The survey by Ul Hassan et al.~\cite{hassan2019differential} provides a thorough overview of DP applications across cyber-physical systems, with a notable emphasis on smart grids. Research in DP for smart grids has primarily focused on three areas: grid demand response, smart building operations, and grid data collection with fog computing. In demand response, DP methods like data masking using Laplacian noise have been explored to protect consumer data without compromising utility operations~\cite{barbosa2016technique}. For smart buildings, which are integral to urban development, DP is used to secure sensor data streams and Internet traffic, ensuring the privacy of inhabitants against potential intrusions \cite{chen2017pegasus,liu2018epic,pappachan2017towards,eckhoff2017privacy,jia2017privacy,ghayyur2018iot}. DP Smart meter load monitoring has been studied in~\cite{jawurek2012sok}. Integrating DP with fog computing has shown the potential to enhance privacy and operational efficiency in smart grids. This approach safeguards data during transmission and storage in fog nodes, protecting against privacy breaches without significant impacts on system performance \cite{xu2018distilling,cao2019achieving}. The employment of DP in demand-data reporting has been shown to preserve the confidentiality of individual consumption patterns while still providing aggregated data useful for grid optimization~\cite{tran2022smart}. Further applications of DP in clustering load profiles have enabled the grouping of similar consumption patterns without infringing on the privacy of individual users~\cite{ravi2022differentially}. This same study illustrates how DP techniques can facilitate the sharing of load profiles for research purposes, effectively balancing the need for data utility with privacy considerations.}

\added{Despite progress, securing grid users' data remains a complex issue, with DP providing a promising path forward across various scenarios. However, areas like fault information transmission, load profiling, and billing information privacy still demand attention to achieve comprehensive privacy protection in smart grid applications.}

\subsubsection{\added{DP Bad Data Detection in Smart Grids}}\label{sec:lit_review_dp_bdd}
\added{A handful of studies have explored the application of DP for bad data and FDI attack detection within smart grids, addressing the critical balance between privacy protection, system security, and data utility. Hossain et al.~\cite{hossain2021privacy} delve into the dual role of DP in smart grids, noting its capacity to safeguard user privacy while potentially enabling integrity attacks through privacy-preserving noise. They propose a tailored DP design strategy focused on mitigating the effects of FDI attacks. Their work extends to assessing the viability of DP in smart grid environments, especially under adversarial conditions, and evaluates the implications on the quality of service. Specifically, they analyze the sum query on a database of measurements from a $\mu$PMU dataset with the addition of simple Laplacian noise, from which they derive optimal strategies for both attack and defense scenarios. Gaboardi et al.~\cite{gaboardi2016differentially} introduce a novel approach to conducting chi-squared tests for goodness of fit and independence that adhere to DP constraints. Their method is innovative in that it modifies classical statistical tests to incorporate DP mechanisms, thus ensuring the privacy of sensitive data. The study presents both Monte Carlo-based and asymptotically aligned tests that adjust for DP-induced noise, highlighting a methodological advancement in integrating privacy preservation within statistical analysis. Lin, et. al.~\cite{lin2024privacy} present a federated learning-based algorithm for distributed and privacy-preserving FDI attack detection that allows state owners to collaboratively generate a global detection model without extensive data transmission, thus protecting data privacy by integrating artificial Gaussian noise into the local model estimations.}

\subsection{Contributions}%
\added{Building on recent advancements and addressing enduring challenges in cyber-defense, our work specifically focuses on enabling bad data detection by entities that may not be fully trusted, all while preserving the privacy of critical system data. This focus marks a distinct departure from existing approaches that primarily enhance the detection mechanisms of FDI attacks or apply DP in a general context. Our contributions, tailored to this unique challenge within the realm of power systems' security and privacy, include:}
\begin{itemize}
    \item \added{A novel chi-squared noise DP mechanism that enhances privacy in querying grid measurements for detecting bad data and anomalies. Applied to the norm of the residual error of the power systems' state estimate~\cite{monticelli2000electric,zhao2019power,liu2011false}, this mechanism is versatile enough to be used for any quadratic queries following a chi-square distribution. This approach enables bad data detection by third parties without compromising the confidentiality of system states or matrices.}
    \item \added{An approximation of the chi-squared mechanism to a Gaussian mechanism for stochastic queries in large systems, optimizing the balance between privacy preservation and analytical utility.}
\end{itemize}

\begin{figure}[!htbp]
  \centering
  \includegraphics[width=0.4\textwidth]{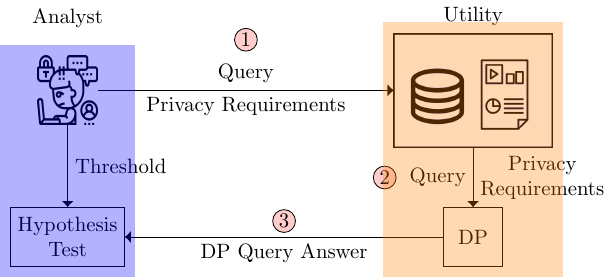}
  \caption{Illustration of our proposed DP mechanism for FDA detection.}
  \label{fig:dp_schematic}
\end{figure}

\added{In contrast to existing approaches detailed in \Cref{sec:lit_review_dp_bdd}, our methodology eschews the direct application of DP noise to measurements in favor of targeting the residual, as formalized in \Cref{sec:wls}. This strategy not only simplifies analytical processes but also demands lower privacy budgets to effectively protect system data. Furthermore, while prior studies consider deterministic and static measurement models, our framework innovatively accounts for the stochastic nature of measurements. By focusing on ensemble-based privacy rather than individual measurement instantiations, we offer a unique contribution to the DP landscape. This methodological innovation permits the precise tailoring of a chi-square DP mechanism for quadratic queries, marking a notable advancement in privacy-preserving techniques for power systems. Our approach underscores the utility of privacy-preserving techniques in supporting collective defense strategies, filling gaps in information sharing and technology, and promoting a more resilient grid against cyber threats in DER-rich environments.}
\deleted{While these initiatives demonstrate the efficacy of DP in mitigating privacy concerns, our work extends the application of DP to the critical task of detecting false data injection attacks, presenting a novel mechanism tailored for preserving privacy in the context of anomaly detection within power systems. To address the research gap and to overcome the threat of FDIAs (described in detail in \ref{sec:threat_model}, this paper describes a novel DP chi-square noise mechanism enabling third-party detection of possible attacks, without revealing private information about power injections or the system matrix.} An illustration of our proposed mechanism is shown in \cref{fig:dp_schematic}.

The remainder of the paper is organized as follows. \Cref{sec:problem_statement} defines the measurement model, performs preliminary analysis on the least squares residual of the state estimation problem, and introduces the threat model. In \Cref{sec:proposed_methodology}, we first introduce the definitions related to DP before presenting our novel DP mechanism and its Gaussian approximation for sharing the residuals with third parties. \Cref{sec:results} showcases the numerical results, and finally, \Cref{sec:conclusion} concludes the paper.

{\bf Notation:} Boldfaced lowercase (uppercase, respectively) letters denote vectors (matrices, respectively), and $x_i$ ($X_{ij}$, respectively) denotes the $i$\textsuperscript{th} element of vector $\bm{x}$ (the $ij$\textsuperscript{th} entry of matrix $\bm{X}$, respectively). Calligraphic letters denote sets, and $|\cdot|$ represents the cardinality of a set. Furthermore, $[N]$ denotes the set of integers ${0, 1, \ldots, N-1}$.

\section{\added{Preliminaries, }Threat Model and Problem Statement}\label{sec:problem_statement}
In power systems operations, state estimation algorithms are used to fit the observed measurements collected from the system and make informed decisions. State estimation algorithms need to be robust to a variety of errors arising from measurement errors, modeling errors, uncertainty in the model parameters, and bad (maliciously placed or otherwise) data. In this paper, we are motivated by the problem of bad data injection attacks on the observed measurements, although the method applies to detecting other anomalies. Traditionally, the analysis of residuals in state estimation has been utilized to detect the presence of bad data: data is considered ``bad'' if the error with respect to the model is higher than what is statistically consistent with the measurement noise. 
%it does not accurately represent the actual physical state of the power system or if it is intentionally manipulated to deceive the state estimation algorithm. 
%Bad data can disrupt the accuracy and reliability of the state estimation process, leading to 
Incorrect estimation of system states can compromise operations and control decisions. 
In this section, we briefly introduce the measurement model and the residual-based bad data detection (BDD) algorithm.

\subsection{Measurement Model and False Data Attack}\label{sec:bdd}
The measurement model relates the observed measurements $\bm{z}$ to the system states $\bm{x}$ and the noise; for additive noise such models can be expressed as:
\begin{equation}
  \bm{z} = \bm{h}(\bm{x}_o) + \bm{\eta} + \bm{a}.\label{eq:meas_model}
\end{equation}
Here, $\bm{z} \added{\in \mathbb{R}^m}$ represents a vector of observed grid measurements, which can include various quantities such as bus injections, bus voltages, and also possibly the so-called pseudo-measurements. The \textit{ground-truth} system states, denoted by $\bm{x}_o \added{ \in \mathbb{R}^n}$, correspond to the voltages at different buses in the power system. 
The function $\bm{h}$ reflects the physical model that ties the state to the quantity measured in the noiseless case and depends on the power system parameters, including the properties of the lines and transformers.
The measurement noise is captured by the random vector $\bm{\eta} \added{\in\mathbb{R}^m}$, which is assumed to follow a Gaussian distribution with mean $\bm{0}$ and covariance $\sigma^2\bm{I}$\footnote{This is without loss of generality since it is always possible to pre-whiten the noise.}.
Additionally, the vector $\bm{a}$ represents the deterministic sparse vector of bad data injections. The sparsity of $\bm{a}$ implies that only a subset of the measurements is targeted by the adversary. If a particular meter is affected by an adversary, the corresponding entry $a_i$ in $\bm{a}$ is non-zero.%, indicating the presence of a bad data injection at that sensor. 

It is important to note that any other errors arising from modeling and uncertainty in model parameters are combined with the measurement errors $\bm{\eta}$ and assumed to be independent of the system parameters. By considering this measurement model, we investigate the effects of bad data injections on the observed measurements and describe one of the widely used methods to detect such attacks.

% Before describing the hypothesis testing framework to detect FDAs, we make the following comment on the validity of the linear approximation considered above:
\begin{remark}\label{rem:meas}
    The measurement model in~\cref{eq:meas_model} is exactly linear for Phasor Measurement Units (PMUs) whose model is:
    \begin{equation}
    \bm{z} = \begin{bmatrix}
               \bm{v} \\
               \bm{i}
             \end{bmatrix} = \begin{bmatrix}
                               \bm{I} \\
                               \bm{Y}
                             \end{bmatrix}\bm{v} + \bm{\eta},\label{eq:pmu_meas}
    \end{equation}
    where $\bm{v}$ is the vector of voltages at the grid buses, $\bm{i}$ is the vector of corresponding currents, and $\bm{Y}$ is the admittance matrix. In this linear model, the voltages vector is also the state $\bm{x}_o$ and the function $\bm{h}$ can be expressed as $\bm{h}(\bm{x}) = \bm{H}\bm{x}$, where $\bm{H}$ is the matrix shown in \cref{eq:pmu_meas}. Any linearized power flow model, such as those proposed in~\cite{baran1989optimal, baran1989network}, including the DC power flow models, are also special cases.
    
    For the non-linear AC power flow model, the analysis relies on a first-order approximation of the measurement model, substituting the Jacobian matrix (refer to \Cref{app:non-lin}) with the system matrix $\bm{H}$, as we will discuss in the next section.
\end{remark}

\subsection{\replaced{BDD}{FDA detection} via Weighted Least Squares}\label{sec:wls}
In this section, we will review classical bad data outlier detection methods to define test statistics that can be shared to determine if the system is experiencing an anomaly, without directly sharing the measurements $\bm{z}$, which could potentially reveal system and state information.

The Weighted Least Squares (WLS) method is commonly used to estimate the system state by minimizing the weighted sum of squared residuals (WSSR), where the weights are determined by the inverse of the covariance matrix. In the case where the observations are pre-whitened, as assumed in \cref{eq:meas_model}, we can consider the equivalent Least Squares (LS) problem without loss of generality:
\begin{equation}
  \bm{x}^{\star} = \argmin_{\bm{x}} \sigma^{-2}\|\bm{z} - \bm{h}(\bm{x})\|^2,\label{eq:wls}
\end{equation}
where $\bm{x}^{\star}$ is the state estimate given measurements $\bm{z}$ that follow the measurement model in \cref{eq:meas_model}, and $\bm{x}$ is the optimization variable. 

Next, \replaced{for convenience's sake, we analyze the linear measurement model but show in \Cref{app:non-lin} that the non-linear model can be linearized by utilizing the Jacobian matrices computed at the current state.}{we will first analyze the linear measurement model case, and then show how to extend the analysis to the non-linear case.} For the linear case, the WSSR can be written as:
\begin{align}
  \qz &= \sigma^{-2}\|\bm{z} - \bm{H}\bm{x}^{\star}\|^2 %= \sigma^{-2}\|\bm{z} - \bm{H}\left(\bm{H}^T\bm{H}\right)^{-1}\bm{H}^T\bm{z}\|^2\nonumber\\
  := \sigma^{-2} \|\bm{P}(\bm{\eta} + \bm{a})\|^2,\label{eq:wssr_lin}
\end{align}
where we use that $\bm{x}^{\star} = \left(\bm{H}^T\bm{H}\right)^{-1}\bm{H}^T\bm{z}$ and set $\bm{P} := \bm{I} - \bm{H}\left(\bm{H}^T\bm{H}\right)^{-1}\bm{H}^T = \bm{P}^2$ as the orthogonal projection (or hat) matrix. It is used to project the observed measurements onto the space spanned by the columns of the Jacobian matrix. In the context of detecting bad data, the matrix $\bm{P}$ is utilized to compute the weighted sum of squared residuals (WSSR) and plays a crucial role in determining the statistical properties of the residual test statistic.

We formulate the detection of bad data as a hypothesis-testing problem with two hypotheses. The null hypothesis $\Hza$ represents the absence of an attack, where $\bm{a}=\bm{0}$. The alternative hypothesis $\Hoa$ corresponds to the presence of an attack, indicating that $\bm{a}\neq \bm{0}$. We write this formally as:
\begin{subequations}
  \begin{align}
    \Hza & : \bm{a} = \bm{0} \qquad(\text{no attack}) \\
    \Hoa & : \bm{a} \ne \bm{0} \qquad(\text{attack})
  \end{align}
\end{subequations}
The query to perform the hypothesis test is the residual test statistic $\qz$ in \cref{eq:wssr_lin}, which allows the analyst to compare the WSSR to a threshold $\tau$. If the WSSR is below $\tau$, we accept the null hypothesis $\Hza$, indicating that there is no attack. Otherwise, if the WSSR exceeds $\tau$, we reject $\Hza$ and accept the alternative hypothesis $\Hoa$, suggesting the presence of an attack, i.e.:
\begin{equation}
  \qz \mathop{\gtrless}_{\Hza}^{\Hoa} \tau.\label{eq:test}%
\end{equation}
By sharing the WSSR, we allow the analyst to freely choose the threshold $\tau$ and determine the optimal trade-off between the probability of false alarm (accepting $\Hoa$ when $\Hza$ is true) and the probability of detection (correctly accepting $\Hoa$ when $\Hoa$ is true). Both probabilities are influenced by the specific values of the bad data vector $\bm{a}$ and the chosen threshold $\tau$.

Under the assumption of Gaussian additive noise $\bm{\eta}$ for hypothesis $\Hoa$, the WSSR follows a non-central chi-square distribution with $r$ degrees of freedom, where $r$ is the rank of the matrix $\bm{P}$. The WSSR is centered at $\theta^2 = \sigma^{-2}\|\bm{P}\bm{a}\|^2$, which represents the squared norm of the projection of the bad data vector onto the subspace orthogonal to the columns of $\bm{H}$, i.e.:
\begin{equation}
  \qz \sim \chi^2_{r}\left(\sigma^{-2}\|\bm{P}\bm{a}\|^2\right),\label{eq:m>n}
\end{equation}
where $r = \mathrm{rank}(\bm{P}) = m-n$. For the null hypothesis, the WSSR follows a central chi-square distribution with $r$ degrees of freedom.

\begin{remark}\label{rem:stoch}
The stochastic nature of $\qz$ will play a fundamental role in the development of our privacy mechanism, which diverges from conventional differential privacy methods designed for deterministic queries. The details of our mechanism and its privacy considerations will be discussed in the following sections.
\end{remark}

\subsection{Special Case of the Measurement Model}\label{sec:spec_cases}
When $m<n$, the null space is empty, and therefore there are no residuals to share. We propose two approaches.
First, we suggest estimating the system state by employing a regularized weighted least squares (RWLS) objective, given by:
\begin{equation}
  \bm{x}^{\star} = \argmin_{\bm{x}} \left(\sigma^{-2}\|\bm{z} - \bm{h}(\bm{x})\|^2 + \lambda \|\bm{x}\|^2\right),\label{eq:rwls}
\end{equation}
where $\lambda$ is the regularization parameter that for $\bm{h}(\bm{x}) = \bm{H}\bm{x}$ is solved by:
\begin{equation}
  \bm{x}^{\star} = \left(\bm{H}^T\bm{H} + \lambda\sigma^2 \bm{I}\right)^{-1}\bm{H}^T\bm{z},\label{eq:reg_state}
\end{equation}
and the WSSR is given by:
\begin{equation}
  \qz = \frac{\bm{z}\bm{P}_{\lambda}^T\bm{P}_{\lambda}\bm{z}}{\sigma^2} \sim \chi^2_{r}(\theta_{\lambda}^2),\label{eq:m<n}
\end{equation}
where $r$ is the rank of $\bm{P}_{\lambda}$ and
\begin{align}
  \bm{P}_{\lambda} & := \left[\bm{I} - \bm{H}\left(\bm{H}^T\bm{H} + \lambda\sigma^2 \bm{I}\right)^{-1}\bm{H}^T\right],\label{eq:P_lambda} \\
  \theta_{\lambda}^2    & := (\bm{H}\bm{x} + \bm{a})^T\bm{P}_{\lambda}^2(\bm{H}\bm{x} + \bm{a})/\sigma^2.\label{eq:mu_lambda}
\end{align}
Note that \cref{eq:reg_state} reduces to the ordinary least squares solution of $\bm{x}^{\star} = \left(\bm{H}^T\bm{H}\right)^{-1}\bm{H}^T\bm{z}$ when the measurement model contains redundant measurements, i.e., when $m > n$, and by setting $\lambda = 0$.

The second option is an alternative form of regularization based on Graph Signal Processing (GSP)~\cite{ramakrishna2019detection}. In the context of a given system, the phasors vector $\bm{v}$ can be regarded as \textit{low-pass graph signal}~\cite{ramakrishna2019detection}. This implies that their empirical covariance matrix has dominant components in the space spanned by the $\kappa < n$ least significant eigenvectors of the system matrix $\bm{Y}$. In other words, we can approximate $\bm{v}$ as $\bm{v} \approx \bm{U}_{\kappa} \tilde{\bm{v}}_{\kappa}$, where $\bm{U}_{\kappa}$ is an $n \times \kappa$ matrix.
We can update the linear model as $\bm{z} = \bm{H}'\bm{x}_o' + \bm{\eta} + \bm{a}$, where $\bm{H}' = \bm{H} \bm{U}_{\kappa}$ and $\bm{x}_o' = \tilde{\bm{v}}_{\kappa}$. The LS solution can still be applied as long as $\kappa < m$. Even when $m > n$, this method is useful because it can handle stealth attacks (see Remark \ref{rem:stealth} in the next subsection). However, it should be noted that the residual in this case may reveal system information, as clarified next.

\subsection{Threat Model}\label{sec:threat_model}
Depending on $m$, $n$, and $r$, if the WSSR is published to an analyst, issues relating to the disclosure of the state (only when $m < n$) and the system matrix may arise. This can be observed in \cref{eq:m>n} and \cref{eq:m<n} where the WSSR depends on the system matrix in the former and both the system matrix and the system state in the latter. The privacy leakage is summarized in \Cref{table:leakage}:
\begin{table}[h!]
  \centering
  \begin{tabular}{ccc}
    \toprule
    System Matrix Size                                      & System Matrix                   & System State                      \\
    \midrule
    $m>n$ & Disclosed & Secure \\
    \midrule
    $m<n$  & Disclosed & Disclosed   \\
    \bottomrule
  \end{tabular}
  \caption{Privacy protection of the system state and matrix.}
  \label{table:leakage}
\end{table}

The publication of system matrices or system states to third parties is a threat to the security and resilience of the electric grid system as they provide valuable information about the system's vulnerabilities. 
%The electric grid system is increasingly interconnected and dependent on digital technologies, making it vulnerable to cyber threats, including data breaches, malware attacks, and ransomware attacks. 
Publishing information about the system's topology, load distribution, or power flow patterns can help attackers identify critical infrastructure components. This reconnaissance information can then be used to plan targeted attacks (such as power outages, equipment damage, or even physical attacks on infrastructure components) that result in severe consequences, including service disruptions and financial losses.
 %The publication of electric grid system matrix or system state can provide attackers with valuable information about the system's architecture and protocols, allowing them to identify and exploit vulnerabilities.
%For example, publishing the electric grid system's topology can reveal critical infrastructure components, such as generators or substations, which can be targeted with malware or ransomware attacks. 
Attackers can also use the system's state information to identify vulnerabilities in the system's control systems, such as SCADA systems or energy management systems, and exploit them for cyberattacks.
%Moreover, the grid is a complex and interdependent network of components, including generators, transformers, transmission lines, and distribution networks. Any disruption or compromise to any of these components can have a ripple effect on the entire system, potentially causing cascading failures and blackouts. 

The publication of system matrices or system states can also pose a privacy risk to energy consumers. The grid system collects and analyzes vast amounts of data on energy consumption patterns, which can be used to infer a consumer's daily routines, lifestyle, and even location. Such information can be exploited for social engineering attacks, such as phishing or spear-phishing, or other forms of cybercrime.
Moreover, the privacy of energy consumers is a fundamental right, and any compromise to this right can erode public trust in the operators. 
%Therefore, it is essential to protect the privacy of energy consumers and ensure that any disclosure of the electric grid system matrix or system state does not compromise this right.
For all the aforementioned reasons the system matrix or system states must remain confidential.

As discussed in the prior sections, traditional rules of thumb adopted by specific industries have flawed or no quantification of privacy guarantees, and anonymization often fails in the presence of substantial side information. In this paper, we address the threat posed by a third-party analyst who may be able to deduce the system state $\bm{x}$ or the system matrix $\bm{H}$ by analyzing the residual query~\added{\cite{lin2024privacy}}.

\begin{remark}[Internal and External Threats]
We do not consider insiders (of the organization that stores the data) with legitimately acquired access to the data as threats. 
Instead, we are concerned with the inference of a data point's involvement after a particular aggregate query has been published to an external, untrustworthy third party.
\end{remark}

\begin{remark}[Stealth Attacks]\label{rem:stealth}
An additional area of concern is stealth attacks where the attacker injects a sparse vector. Here, non-zero entries of the attack vector that correspond to the sensors being attacked are modeled such that residual in \cref{eq:minWSSR} is unaffected even with the perturbed state:
\begin{equation}
    \bm{P}\bm{a} = \bm{0}
\end{equation}
Here, the attacker can alter the algorithm's output without any change in the loss function of the state estimation problem \cref{eq:wls}. These types of attacks are only possible when a malicious agent possesses complete knowledge about the system and a non-trivial null space of $\bm{P}$ exists. Detecting and mitigating such attacks is challenging, particularly in the absence of a specially imposed structure on the actual measurement vectors. The literature proposes various methodologies to detect stealth attacks. As mentioned in the previous section, we recommend using the GSP-based method presented in \cite{ramakrishna2019detection}. This approach aligns well with our measurement model description and is highly effective in detecting stealth attacks. 
%The authors of the paper employ a GSP approach to investigate FDIAs on electric power systems' synchrophasor measurements. By modeling PMU data as low-pass graph signals, they explore the utilization of this structure to develop enhanced BDD algorithms capable of effectively detecting FDI attack signatures. Incidentally, the authors also employ a hypothesis test in their approach, where the test statistic has the same properties as $\qz$, although with different a center. 
\end{remark}

In the next section, after formally defining the concept of DP, we introduce our proposed DP mechanism for sharing the test statistic. This mechanism aims to safeguard the differential privacy of both the system matrix and the state.

\section{Differentially Private Bad Data Detection}\label{sec:proposed_methodology}
Motivated by overcoming the disclosure issues of BDD algorithms, in this section, we describe a novel methodology for the publication of a differentially privatized test statistic $\qz$ and show how to adjust the performance guarantees of the hypothesis test to account for the loss in accuracy due to the DP mechanism. Before describing our novel methodology, we first provide a brief description of differential privacy.

\subsection{ Preliminaries}\label{sec:dp_prelims}
In the context of a dataset $\bm{z}$ and a query $\mathbb{q}$, we use the notation $\tilde{\mathbb{q}}(\bm{z})$ to represent the differential private answer to the query $\mathbb{q}$. The random outcome of the query, post the application of the DP mechanism, is denoted as $\qtilde$, which belongs to the set $\mathcal{Q}$ and follows a distribution $f(\qtilde|\bm{z})$. This distribution is a probability density function for continuous random queries or a probability mass function for discrete random variables.
The common definition of differential privacy from~\cite{dwork2006calibrating,dwork2006our} is:
\begin{defn}[$(\epsilon,\delta)$-Differential privacy]
  A randomized mechanism $\tilde{\mathbb{q}}$ is $(\epsilon,\delta)$-differentially private if for all neighboring datasets $\bm{z}$ and $\bm{z}'$ that differ in one point, for any arbitrary event pertaining to the outcome of the query, the randomized mechanism satisfies the following inequality
  \begin{equation}\label{eq:e-dp}
    \forall \mathcal{S},\quad
    Pr(\tilde{\mathbb{q}}(\bm{z}) \in \mathcal{S}) \leq \exp(\epsilon)Pr(\tilde{\mathbb{q}}(\bm{z}') \in \mathcal{S}) + \delta,
  \end{equation}
  where $Pr(\mathcal{A})$ denotes the probability of the event $\mathcal{A}$, for some privacy budget $\epsilon\geq 0$ and $\delta \in [0,1]$.
\end{defn}
Note that, since $\delta$ is a bound that may not be tight, smaller values of $\delta$ are possible. Hence, $(\epsilon,\delta)$ guarantees are sufficient but not necessary conditions. A second definition in terms of the privacy leakage function is:
\begin{defn}[$(\epsilon,\delta)$-Probabilistic Differential privacy]\label{def:probabilisticDP} The so-called privacy leakage function  $L_{\XX}$ is the log-likelihood ratio between the two hypotheses that the query outcome $\qtilde$ is the answer generated by the data $\bm{z}$ or the data $\bm{z}'$ that differ by one element. Mathematically:
  \begin{equation}
    L_{\XX}(\qtilde):=\log \frac{f(\qtilde|\bm{z})}{f(\qtilde|\bm{z}')},
  \end{equation}
  A randomized mechanism $\tilde{\mathbb{q}}(\bm{z})$ is $(\epsilon,\delta)$ differentially private for $\bm{z}$ if and only if:
  \begin{equation}\label{eq:def1}
    \sup_{\XX}~Pr\left(|L_{\XX}(\qtilde)| \leq \epsilon\right) \geq 1 - \delta.
  \end{equation}
\end{defn}
It can be shown that $(\epsilon,\delta)$-PDP is a strictly stronger condition than $(\epsilon,\delta)$-DP \cite{mcclure2015relaxations}.

\subsection{DP for Stochastic Queries}
Earlier in this section, we provided an overview of two definitions of DP. However, we highlight an important caveat regarding these definitions -- they primarily focus on protecting the DP of individual elements, denoted as $z$, within a database $\bm{z}$. They aim to conceal the presence or absence of each element in $\bm{z}$. Traditionally, mechanisms derived based on these definitions assume that the database is deterministic, lacking any stochastic aspects in its entries. For instance, consider an averaging query on a database consisting of the income of a group of people or biographical details of individuals surveyed for a census~\cite{garfinkel2022Differential}.

However, in this paper, we address measurement vectors $\bm{z}$ that arise from a stochastic measurement model based on the physics of the electric grid, as discussed in \Cref{sec:wls} and \Cref{rem:stoch}. In this scenario, our primary focus is on safeguarding the DP of the system configuration that generates the measurement model, rather than focusing on an individual instantiation of its measurements. This approach was motivated by the reasons detailed in \Cref{sec:threat_model}. It is worth noting that the data owner possesses knowledge of the system configuration only in scenarios where the electric utility itself is the data owner. In all other cases, neither the data owner nor the external analyst has access to the system configuration.

For instance, suppose we aim to protect the DP of the elements in the matrix $\bm{H}$. In this case, each system configuration gives rise to an \textit{ensemble of query instantiations}. By considering the stochastic nature of the measurements and focusing on the DP of the system rather than individual measurements, our paper introduces a novel perspective in the realm of differential privacy mechanisms.

The traditional definitions of DP include a neighboring database at distance one (or, in other words, differing in one element). Similarly, we define a distance one neighbor to the system matrix as follows:
\begin{defn}[Distance one neighborhood]\label{def:neighbor}
    Consider a system matrix $\bm{H} \in \mathbb{R}^{m \times n}$. The distance one neighborhood of $\bm{H}$ is defined as the set of all matrices that differ from $\bm{H}$ in exactly one row. More formally,
    \begin{equation}
        \left\{ \bm{H}' \in \mathbb{R}^{m\times n}\mid \bm{H}' = \bm{H} + \bm{e}\bm{\Delta}_h^T\right\},
    \end{equation}
    where $\bm{H}$ and $\bm{H}'$ differ in the $m$\textsuperscript{th} row (without loss of generality), $\bm{e} \in \bm{R}^m$ is a coordinate vector with its $m$\textsuperscript{th} entry set to $1$ and all other entries set to $0$. Additionally, $\bm{\Delta}_h:=\left(\bm{h}' - \bm{h}\right)$, where $\bm{h}$ and $\bm{h}'$ represent the $m$\textsuperscript{th} rows of $\bm{H}$ and $\bm{H}'$, respectively.
\end{defn}
Consequently, when considering a matrix $\bm{H}'$ from the distance one neighborhood of $\bm{H}$, the corresponding vector of measurements $\bm{z}'$ differs from $\bm{z}$ in exactly one element:
\begin{equation}
z'_i = \begin{cases}
         z_i, & i < m, \\
         \bm{h}'^T\bm{x} + a + \eta,  & i = m
       \end{cases} ~ \Rightarrow ~
\bm{z}' = \bm{H}'\bm{x} + \bm{a} + \bm{\eta}.\label{eq:z_differ}
\end{equation}
In turn, the distance one neighborhood WSSR is given by: 
\begin{equation}
\mathbbm{q}(\bm{z}') = \sigma^{-2}(\bm{a} + \bm{\eta})^T\bm{P}'(\bm{a} + \bm{\eta}),
\end{equation}
where $\bm{P}' = \bm{I} - \bm{H}'\bm{H}'^\dagger$. This implies that the residual follows the non-central chi-square distribution with $r:=m-n$ degrees of freedom and non-centrality parameter $\theta'^2 := \sigma^{-2}\|\bm{P}'\bm{a}\|^2$:
\begin{equation}
\mathbbm{q}(\bm{z}') \sim \chi^2_{r} (\theta'^2).
\end{equation}

\begin{remark}
    In our framework, it is important to note that the term \textit{differential} in differential privacy arises from the need to conceal whether a measurement is a result of the system configuration of $\bm{H}$ or one of its neighboring configurations $\bm{H}'$. This not only hides the origin of the measurement as part of an ensemble but also enables a more traditional interpretation in terms of the differential of the actual measurement vector, as illustrated in \cref{eq:z_differ}. 
\end{remark}

Finally, to derive a differentially private mechanism for answering the residual query, we will rely on \Cref{def:probabilisticDP}, which provides a direct statistical interpretation. As $(\epsilon, \delta)$ values approach zero, the log-likelihood ratio, which serves as a sufficient statistic for determining whether the randomized answer is generated from neighboring datasets $\bm{z}$ or $\bm{z}'$, produces mostly incorrect or unreliable outcomes. In other words, there is a non-zero probability of the test yielding incorrect results. This trade-off in terms of answer accuracy needs to be carefully considered.
% \begin{figure*}%
%   \centering%
%   \includegraphics[width=0.75\textwidth]{./figs/schematic/attack}
%   \caption{Illustration of the model.}%
%   \label{fig:model}%
% \end{figure*}%

\subsection{Differentially Private Chi-Squared Noise Mechanism}
As seen in \cref{eq:m>n,eq:m<n}, the WSSR query is a non-central chi-square random variable.
In this section, we propose a novel additive noise DP mechanism where the WSSR is treated with a random noise drawn from the chi-squared distribution as follows:
\begin{equation}
  \tqz = \qz + \nu \quad\text{where}\quad \nu \sim \chi^2_{r'}(0),\label{eq:chi_sq_mech}
\end{equation}
which implies that $\tqz$ is also a non-central chi-square random variable with $\tilde{r} := r+r'$ DoF and centered at $\theta^2$, i.e.:
\begin{equation}
  \tqz \sim \chi^2_{\tilde{r}}(\theta^2).
\end{equation}
With this in mind, we state the following theorem guaranteeing the $(\epsilon, \delta)$-DP of the chi-square noise mechanism with its proof in \cref{app:chi}.

\begin{thm}[Chi-square mechanism is $(\epsilon,\delta)$-DP]\label[thm]{thm:eps-del_chis}
  The mechanism in \cref{eq:chi_sq_mech} is $(\epsilon,\delta)$-DP for all pairs of neighboring measurement sets $\bm{z}$ and $\bm{z}'$ differing in exactly one measurement, where the guarantee $\delta$ is given by:
  \begin{align}
    \delta \!\!&=\!\! \max_{\theta, \theta'}\!\!\left[\mathrm{Q}_{\frac{\tilde{r}}{2}}\!\!\left(\!\!\theta,\frac{\epsilon}{\theta'\!\! -\!\! \theta} \!-\! \frac{\theta' \!\!+\!\! \theta}{2}\right) \!\!+\!\! \mathrm{Q}_{\frac{\tilde{r}}{2}}\left(\theta,\frac{\epsilon}{\theta' \!\!-\!\! \theta} \!\!+\!\! \frac{\theta' \!\!+\!\! \theta}{2}\right)\right],
  \end{align}
  where $\mathrm{Q}_{s}(a,b)$ is the Marcum Q-function of order $s>0$ with $a>0$ and $b\geq 0$.
\end{thm}
The sensitivity analysis is undertaken in \cref{app:sensitivity} and provides explicit expressions for the $\delta$ under assumptions about the system Jacobian.

While this mechanism can be analyzed for smaller $m$, the analytical calculation to derive $\delta$ for larger values of $m$ is not numerically viable, as the Marcum Q-function becomes degenerate in this regime. Thus, in the following section, we provide a Gaussian approximation for the noisy query $\tqz$ that may be used with the Gaussian mechanism for stochastic queries developed in~\cite{ramakrishna2023differential} to release the residual query.

\subsection{Gaussian Approximation}
The residual query, WSSR, follows a non-central chi-square distribution as discussed in \Cref{sec:wls}. We derive a Gaussian approximation of the WSSR using the following theorem, first proved by~\cite[Theorem 1]{zhang2005approximate}. This provides us with a method for dealing with systems with large $m$ values.

\begin{thm}[Gaussian Approximation of $\qz$]\label[thm]{thm:normal_approx}
  Given a measurement vector $\bm{z}$ and the linear measurement model $\bm{z} = \bm{H}\bm{x} + \bm{a} +\bm{\eta}$ with $\bm{\eta}\sim \mathcal{N}(\bm{0},\sigma^2\bm{I})$, and the singular value decomposition of $\bm{H}=\bm{U}\bm{\Sigma}\bm{V}^T$, then the following statements hold:
  \begin{enumerate}[label=(\alph*)]
    \item The WSSR, $\qz$, is a chi-squared-type mixture:
          \begin{equation*}
            \qz = \sum_{i=1}^m d_i z_{U,i}^2, \quad \text{with}\quad z_{U,i}^2 \stackrel{\text{ind}}{\sim} \chi_1^2(\theta_i^2), \quad \forall i \in [m],
          \end{equation*}
          where 
          \begin{align}
            \bm{z}_U &:= \bm{U}^T\bm{z}, \quad \bm{\theta} := \bm{\Sigma}\bm{V}^T\bm{x}+\bm{U}^T\bm{a},\\
            \bm{D} &:= \left(\bm{I} - \bm{\Sigma}\left(\lambda\sigma^2\bm{I}+\bm{\Sigma}^T\bm{\Sigma}\right)^{-1}\bm{\Sigma}^T\right)^2, \text{ and }\\
            \bm{d} &:= \mathrm{diag}(\bm{D}).
          \end{align}
    \item The cumulants of $\qz$ for $\ell = 1,2,\ldots$ are given by:
          \begin{equation}
            \mathcal{K}_\ell(\qz) =  2^{\ell-1} (\ell - 1)  ! \sum_{i=1}^m d_i^{\ell} (1 + \ell \cdot \theta_i^2).
          \end{equation}
          Let $\zeta := \frac{8\mathcal{K}_2^3(\qz)}{\mathcal{K}_3^2(\qz)}$ and $\rho := \max_{1\leq i\leq m} \frac{2 d_i^2\left(1 + 2\theta_i^2\right)}{\mathcal{K}_2(\qz)}$.
          % \begin{equation}
          %   \zeta := \frac{8\mathcal{K}_2^3(\qz)}{\mathcal{K}_3^2(\qz)}, \quad
          %   \rho := \max_{1\leq i\leq m} \frac{2 d_i^2\left(1 + 2\theta_i^2\right)}{\mathcal{K}_2(\qz)}.
          % \end{equation}
          Then, for the normalized $\qz$ given by $\bar{q}(\bm{z}) = \frac{\qz - \mathcal{K}_1(\qz)}{\sqrt{\mathcal{K}_1(\qz)}}$, the following inequality is satisfied:
          \begin{equation}
            \sup_q |f_{\bar{q}(\bm{z})}(q) - \phi(q)| < 0.1323 \left(4 + \frac{0.2503}{(1-8\rho)^2}\right) \zeta^{-\frac{1}{2}}
          \end{equation}
          when $\rho < 1/8$, where $\phi(q)$ is the density function of a standard normal random variable.
    \item If either (i) $\rho \rightarrow 0$ or (ii) $\max_{1\leq i\leq m} \left(1+2\theta_i^2\right) < \infty$ and $\zeta \rightarrow \infty$, is satisfied, then $\bar{q}(\bm{z}) \rightarrow \mathcal{N}(0,1)$.
  \end{enumerate}
\end{thm}
Using \cref{thm:normal_approx}, we can show that:
\begin{align}
  \qz           & \sim \mathcal{N}(\theta_z,\sigma_z^2), \quad \text{where}       \\
  \theta_z      & = \mathrm{Tr}(\bm{D}) + \bm{\theta}^T\bm{D}\bm{\theta}, \quad \text{and} \\
  \sigma_z^2 & = 2\mathrm{Tr}(\bm{D}^2) + 4\bm{\theta}^T\bm{D}^2\bm{\theta}.
\end{align}
The Gaussian DP mechanism that is used in literature is $(\epsilon,\delta)$-DP for a deterministic query. However, in our case, the query is stochastic and, moreover, the variances of the query under two neighboring measurement sets are not the same, i.e., $\sigma_z^2 \ne \sigma_{\X'}^2$. Thus, in the following subsection, we derive the DP guarantees for a stochastic query.

\section{Performance Metrics and Results}\label{sec:results}

In \Cref{sec:dp_prelims}, we presented our chi-square noise mechanism and its Gaussian approximation for publishing residuals of a BDD algorithm. Adding DP noise inevitably corrupts the residual, which will affect the utility of the residual in bad data detection and lead to degraded performance of the hypothesis test in \cref{eq:test}. In this section, the performance of the hypothesis test with the DP noisy residual is quantified through the Receiver Operating Characteristic (ROC) analysis.

The ROC curve is a graphical representation of the trade-off between the probability of detection (denoted by $P_d$ -- it is the probability that a true anomaly is correctly identified as such) and the probability of false alarm (denoted by $P_{fa}$ -- it is the probability that a normal data point is incorrectly identified as an anomaly) of a hypothesis test.

A perfect hypothesis test would have a $P_d$ of 1 and a $P_{fa}$ of 0, which would mean that all true anomalies are identified as anomalies and all normal data points are identified as normal data. However, in practice, no hypothesis test is perfect, so there is always a trade-off between the two.

The area under the ROC curve (AUROC) indicates how well the hypothesis test can distinguish between anomalies and normal data. A higher AUROC indicates that the test is better at distinguishing between anomalies and normal data.

\subsection{Performance of the chi-square noise mechanism}\label{sec:Chi-performance}
% In this section, we discuss the performance of the hypothesis test in terms of its probability of false alarm, $P_{fa}$, and probability of detection, $P_d$. 
In this section, we first derive the probabilities of detection and false alarm for the hypothesis test without any DP noise, that is, with the use of the true residual $\qz$. We then do the same for the residual with the DP noise, $\tqz$.

Suppose the operator sets a probability of false alarm of $\alpha$, then from the definition of the hypothesis test in \cref{eq:test}, we get:
\begin{equation}
  \alpha =: P_{fa} = Pr\left\{ \qz > \tau \mid \mathcal{H}_0 \right\},
\end{equation}
and since $\qz$ is a central chi-square random variable with $r$ degrees of freedom under $\mathcal{H}_0$, we have:
\begin{equation}
  \alpha = 1 - \mathcal{P}\left(\frac{r}{2}, \frac{\tau}{2}\right) := \mathcal{Q}\left(\frac{r}{2}, \frac{\tau}{2}\right),\label{eq:pfa_chi}
\end{equation}
where $\mathcal{P}(\cdot, \cdot)$ is the regularized gamma function and $\mathcal{Q}(\cdot, \cdot)$ is its complementary. Using this relation, we may calculate the threshold to be set as:
\begin{equation}
  \tau = 2\mathcal{Q}^{-1}\left(\alpha, r/2\right).
\end{equation}
Similarly, under $\mathcal{H}_1$, $\qz$ is a non-central central chi-square random variable with a non-centrality parameter of $\theta^2~=~\sigma^{-2}\|\bm{P}\bm{a}\|^2$ and  $r$ degrees of freedom. Then, the probability of detection is given by:
\begin{align}
  P_{d} & = Pr\left\{ \qz > \tau \mid \mathcal{H}_1 \right\} = \mathcal{Q}_{r/2}\left(\sigma^{-1}\|\bm{P}\bm{a}\|, \sqrt{\tau}\right),\nonumber \\
        & = \mathcal{Q}_{r/2}\left(\sigma^{-1}\|\bm{P}\bm{a}\|, \sqrt{2\mathcal{Q}^{-1}\left(\alpha, r/2\right)}\right),\label{eq:pd_chi}
\end{align}
where $\mathcal{Q}_{\{\cdot\}}(\cdot, \cdot)$ is the Marcum Q-function.

In a similar vein, we may compute these probabilities with the DP residual. First, recall that:
\begin{subequations}
\begin{align}
  \mathcal{H}_0 & : \tqz \sim \chi^2_{\tilde{r}}(0) \qquad(\text{no attack})            \\
  \mathcal{H}_1 & : \tqz \sim \chi^2_{\tilde{r}}(\theta^2) \qquad(\text{attack})
\end{align}
\end{subequations}
Suppose $\tilde{r} = r+1$, then, for a threshold of $\tau = 2\mathcal{Q}^{-1}\left(\alpha, r/2\right)$, the probabilities are thus calculated as:
\begin{subequations}\label{eq:DP_pfa_pd_chi}
    \begin{align}
  \tilde{P}_{fa} & \!:=\! Pr\left\{ \tqz \!>\! \tau \!\mid \!\mathcal{H}_0 \right\} \! =\!  \mathcal{Q}\!\left(\frac{\tilde{r}}{2}, \mathcal{Q}^{-1}\left(\alpha, r/2\right)\right),                             \\
  \tilde{P}_{d}  & \!:=\! Pr\left\{ \tqz \!>\! \tau \!\mid \!\mathcal{H}_1 \right\} \! =\!  \mathcal{Q}_{\frac{\tilde{r}}{2}}\!\left(\!\theta, \!\sqrt{2\mathcal{Q}^{-1}\!\left(\alpha, r/2\right)}\right)\nonumber \\
                 & =  \mathcal{Q}_{\frac{\tilde{r}}{2}}\left(\sigma^{-1}\|\bm{P}\bm{a}\|, \sqrt{2\mathcal{Q}^{-1}\left(\alpha, r/2\right)}\right).
\end{align}
\end{subequations}
As shown in \crefrange{eq:pfa_chi}{eq:pd_chi} and \cref{eq:DP_pfa_pd_chi}, the additional degree of freedom required by the DP noise is the main reason for the change in performance. This is because the DP noise increases the variance of the residual, which makes it more difficult to distinguish between normal data and anomalies.

\subsection{Performance of the Gaussian approximation}\label{sec:Gaussian-performance}
A line of analysis is similar to the one undertaken in \cref{sec:Chi-performance} leads to the following false alarm and detection probabilities for the hypothesis test. Recall that $\qz~\sim~\mathcal{N}(\theta_{z},\sigma_{z}^2)$. The moments of the query vary depending on the hypothesis $\mathcal{H}_0$ or $\mathcal{H}_1$. Then let $\theta_{z,h}$ and $\sigma_{z,h}^2$ denote the mean and variance under hypothesis $\mathcal{H}_h$, for $h=0,1$. They are given by:
\begin{align}
  \theta_{z, h}    & = \mathrm{Tr}(\bm{D}) + \bm{\theta}_{h}^T\bm{D}\bm{\theta}_{h},    \\
  \sigma_{z, h} & = 2\mathrm{Tr}(\bm{D}^2) + 4\bm{\theta}_h^T\bm{D}^2\bm{\theta}_h,
\end{align}
where:
\begin{equation}
  \bm{\theta}_h = \bm{\Sigma}\bm{V}^T\bm{x} + h \cdot \bm{U}^T\bm{a} \quad \text{for } h=0,1.
\end{equation}
The false alarm probability is given by:
\begin{equation}
  \alpha =: P_{fa} = \mathcal{Q}\left(\frac{\tau - \theta_{z,0}}{\sigma_{z,0}}\right),
\end{equation}
where $\mathcal{Q}(\cdot)$ is the Gaussian Q function. The threshold can be calculated as:
\begin{equation}
  \tau = \theta_{z,0} + \sigma_{z,0}\mathcal{Q}^{-1}\left(\alpha\right).
\end{equation}
Similarly, the probability of detection is given by:
\begin{equation}
  P_{d} = \mathcal{Q}\left(\frac{\sigma_{z,0}\mathcal{Q}^{-1}\left(\alpha\right) - \Delta_{\theta_z}^{(a)}}{\sigma_{z,1}}\right),
\end{equation}
where $\Delta_{\theta_z}^{(a)} = \theta_{z,1} - \theta_{z,0}$.
Similarly, for the DP residual, we have the following false alarm and detection probabilities:
\begin{align}
  \tilde{P}_{fa} & = \mathcal{Q}\left(\frac{\tau - \tilde{\theta}_{z,0}}{\tilde{\sigma}_{z,0}}\right) = \mathcal{Q}\left(\frac{\tau - (\theta_{z,0} + \theta_{\nu|z})}{\sqrt{\sigma_{z,0}^2 + \sigma_{\nu|z}^2}}\right), \\
  \tilde{P}_{d} & = \mathcal{Q}\left(\frac{\sigma_{z,0}\mathcal{Q}^{-1}\left(\alpha\right) - (\Delta_{\theta_z}^{(a)} + \theta_{\nu|z})}{\sqrt{\sigma_{z,1}^2 + \sigma_{\nu|z}^2}}\right).
\end{align}

\subsection{Numerical Results}
In this section, we provide a comprehensive analysis of the performance of our detection algorithm in the presence of DP noise and compare it with an approach involving input perturbation. We have previously discussed the limitations associated with directly perturbing the measurement vector using input perturbation to protect the system and its state. To further explore this, we consider an input perturbation scenario with $\sigma$ and $\sigma_w$ denoting the standard deviation of the measurement error and DP noise, which is added using the Gaussian mechanism. Recall that in this scenario, the observed measurement vector $\bm{z}$ is given by $\bm{H}\bm{x} + \bm{a} + \bm{\eta} + \bm{w}$.
\begin{figure}[!htbp]
  \centering
  \includegraphics[width=0.22\textwidth]{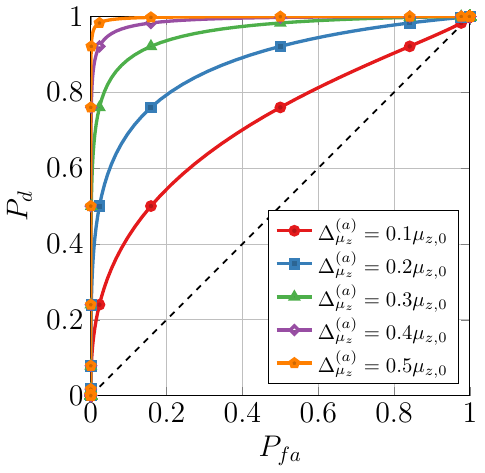}
  \caption{Sensitivity of the test's performance to attack strength.}
  \label{fig:TrueROC}
\end{figure}

Throughout this section, we present a detailed analysis of our detection algorithm's performance under different conditions.
In \cref{fig:TrueROC}, we examine the ROC for binary hypotheses $\mathcal{H}_{\cdot}^{(a)}$ for various values of $\Delta{\theta_z}^{(a)}$. To establish this, we set the means and variances as follows: $\theta_{z,0} = 10$, $\theta_{z,1} = \theta_{z,0}+ \Delta_{\theta_z}^{(a)}$, $\sigma_{z,0} = 1$, $\sigma_{z,1} = 2$. This analysis reveals that the ROC curve deteriorates as the difference $\Delta_{\theta_z}^{(a)}$ between the means of the two hypotheses decreases. This is an expected outcome, as the hypothesis test is more likely to accept the null hypothesis when the difference between the means is small, even under the alternate hypothesis.
\begin{figure}[!htbp]
  \centering
  \includegraphics[width=0.22\textwidth]{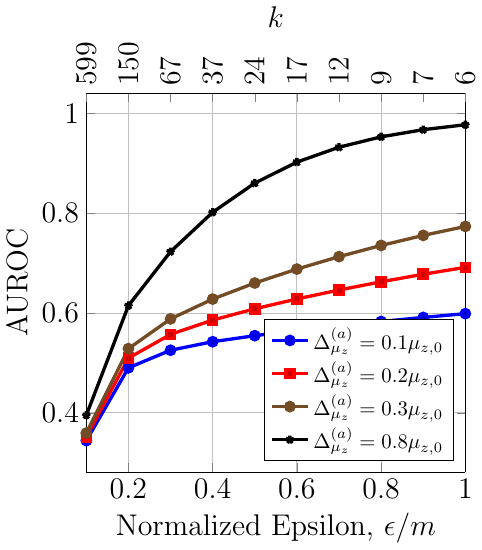}
  \caption{AUROC vs. DP Normalized privacy budget and noise variance scaling ($k$-factor) for input-perturbed measurements $\tilde{\bm{z}} = \bm{z} + \bm{w}$.}
  \label{fig:InputPertROC}
\end{figure}

In \cref{fig:InputPertROC}, we focus on the scenario where input perturbation of the measurement vector $\bm{z}$ is performed before conducting the hypothesis test. We employ the standard Gaussian mechanism with $\delta=0.1$ and a sensitivity set to 1. This analysis is conducted for different values of $\Delta_{\theta_z}^{(a)}$ to understand how varying levels of DP noise impact the algorithm's performance when using input perturbation. We plot the AUROC against the DP privacy parameter $\epsilon$ and the corresponding $k$-factor, where the DP noise variance is denoted as $\sigma_w^2 = k\sigma^2$. Notably, we present this information in terms of the normalized (or the per-element) privacy budget, $\epsilon_o := \epsilon / m$, as we are adding noise to each of the $m$ elements in the $\bm{z}$ vector. As expected, we observe that the AUROC increases with an increase in the per-element privacy budget. This implies that when a smaller standard deviation is used for input perturbation noise, a higher level of performance can be achieved. In practical terms, this figure helps analysts understand the trade-off between the desired level of performance (as defined by the AUROC), the $\Delta_{\theta_z}^{(a)}$, and the allocated privacy budget ($\epsilon$). It provides valuable insights into the resources required to ensure a specific level of performance while safeguarding sensitive information.
\begin{figure}[!htbp]
  \centering
  \includegraphics[width=0.22\textwidth]{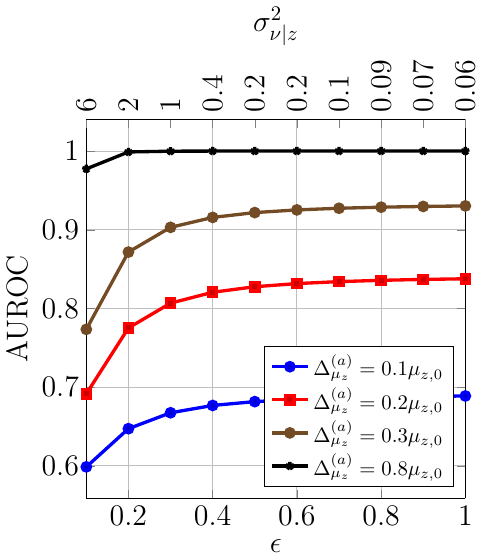}
  \caption{The AUROC of the hypothesis test for different values of $\sigma_{\nu|z}$ when using our mechanism.}
  \label{fig:AUROC}
\end{figure}

In \cref{fig:AUROC}, we illustrate the AUROC of our detection mechanism, incorporating our novel approximate Gaussian DP noise mechanism, across a range of values for $\Delta_{\theta_z}^{(a)}$. Specifically, we set the means and variances as follows: $\theta_{z,0} = 10$, $\theta_{z,1} = 1.3\theta_{z,0}$, $\sigma_{z,0} = 1$, and $\sigma_{z,1} = 4$. It is worth emphasizing that our approach demonstrates superior performance in terms of the privacy budget when compared to input perturbation. This improved efficiency results from our targeted privacy-preserving strategy, which focuses on perturbing the residual query rather than applying noise to the entire measurement vector. As a consequence, we achieve the desired level of performance while minimizing the expenditure of the privacy budget. It's important to note that in this figure, the AUROC curve is plotted directly against the privacy budget, rather than the normalized privacy budget, further underscoring the efficiency of our mechanism.
\begin{figure}[!htbp]
  \centering
  \includegraphics[width=0.3\textwidth]{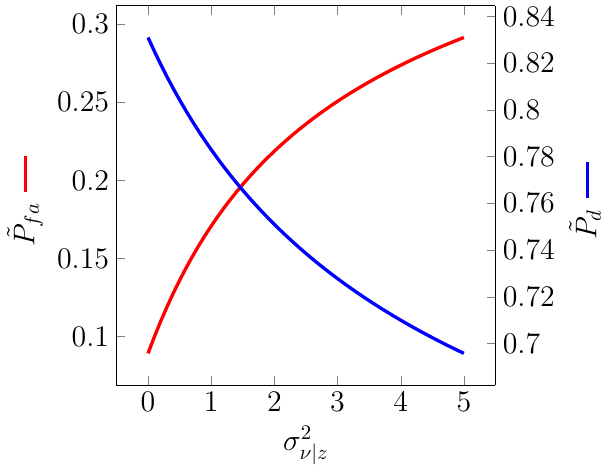}
  \caption{The probabilities of detection and false alarm of the hypothesis test for different values of $\sigma_{\nu|z}$ when a $P_{fa}$ of $0.05$ is required.}
  \label{fig:metrics}
\end{figure}

In \cref{fig:metrics}, we investigate the performance of the hypothesis test across various values of $\sigma_{\nu|z}$ while maintaining a required $P_{fa}$ of 0.05. As with \cref{fig:AUROC}, the means and variances are set as follows: $\theta_{z,0} = 10$, $\theta_{z,1} = 1.3\theta_{z,0}$, $\sigma_{z,0} = 1$, and $\sigma_{z,1} = 4$.
This analysis provides insights into how our algorithm behaves under the constraints of a controlled false alarm rate while introducing varying levels of noise (expressed by different $\sigma_{\nu|z}$ values) to the system. As expected, we observe that the probability of false alarm ($P_{fa}$) increases, and the detection probability ($P_d$) decreases as we introduce noise with increasing variances. This figure serves to highlight the trade-off between algorithm performance, as defined by $P_{fa}$ and $P_d$, and the noise variance, and by extension, the privacy budget allocation.

\section{Conclusion}\label{sec:conclusion}
\replaced{This paper presents a novel DP chi-squared noise mechanism tailored for power systems, emphasizing residual analysis over direct measurement perturbation. This approach simplifies analytics, requires lower privacy budgets, and introduces a mechanism that considers the stochastic nature of power system measurements—a distinct contribution to DP applications. Our methodology enables precise chi-square DP mechanism application to quadratic queries, enhancing privacy-preserving capabilities within power systems and beyond. By focusing on ensemble-based privacy, this work supports the detection of attacks by third parties without exposing system states or matrices, thus bridging significant gaps in cyber defense. Additionally, our approximation of the chi-squared mechanism to a Gaussian mechanism for stochastic queries illustrates the method's adaptability. We advocate for a collective defense strategy, leveraging a distributed detection framework that reduces data transmission and boosts privacy, addressing the need for cooperative cybersecurity solutions in the increasingly interconnected power grid landscape.}
{This paper emphasizes the criticality of accurate measurement data for efficient power system management while addressing the vulnerability of power systems to false data attacks (FDAs). FDAs can lead to severe consequences, including blackouts, equipment damage, and financial losses, affecting millions of people. These attacks can occur at different stages of the power system and target specific components or aim to disrupt the entire system. 
Collaboration and information sharing among utilities are essential for effective FDA prevention and detection. However, the fragmented nature of power grid management and data sensitivity concerns hinder seamless data exchange. To overcome these challenges, the paper proposes the rigorous application of DP mechanisms and metrics.
DP mechanisms provide provable privacy and accuracy trade-offs, protecting sensitive information while preserving data utility. 
To address the research gap, the paper proposes a novel DP chi-square noise mechanism for third-party FDA detection without revealing private information. The proposed mechanism, along with its Gaussian approximation for sharing residuals, is presented in detail. Numerical results showcase the effectiveness and practicality of the proposed DP mechanism.
By embracing DP, stakeholders in the power system sector can strike a balance between data utility and privacy preservation, fostering collaboration, developing effective defense mechanisms, and ensuring regulatory compliance. This approach strengthens the security and reliability of power systems, benefiting millions of electricity users.}

\bibliography{ref}
\bibliographystyle{IEEEtran}

\begin{appendices}

\section{Linearization of the Non-linear Measurement Model}\label{app:non-lin}
This is the case in which $\bm{z}$ contains power flow measurements and, thus $\bm{h}(\bm{x}_o)$ are the AC power flow equations. Here, the WSSR may be written as: 
\begin{align}
    \qz &:= \sigma^{-2}\|\bm{z} - \bm{h}(\bm{x}^{\star})\|^2 = \sigma^{-2}\|\bm{h}(\bm{x}_o) \!+ \bm{\eta} \!+ \bm{a} \!- \bm{h}(\bm{x}^{\star})\|^2\nonumber\\
    &\simeq \sigma^{-2}\|\bm{H} \left(\bm{x}_o - \bm{x}^{\star}\right) + \bm{\eta} + \bm{a}\|^2, \label{eq:wssr}
\end{align}
where the approximation relies on the assumption that $\|\bm{x}_o - \bm{x}^{\star}\|$ is small and the Taylor expansion of $\bm{h}$ around $\bm{x}^{\star}$:
\begin{equation}
    \bm{h}(\bm{x}_o) \simeq \bm{h}(\bm{x}^{\star}) + \bm{H} \left(\bm{x}_o - \bm{x}^{\star}\right),\label{eq:taylor}
\end{equation}
where $\bm{H} = \frac{\mathrm{d}\bm{h}}{\mathrm{d\bm{x}}}$ is the system Jacobian matrix\footnote{We abuse the notation to denote both the Jacobian of $\bm{h}$ and the linear measurement model's system matrix by the symbol $\bm{H}$.}.
Also, at the minimizer, $\bm{x}^{\star}$, the gradient of the objective in \cref{eq:wls} is zero:
\begin{align}
&\bm{0} = \bm{\sigma}^{-2}\bm{H}^T\!\left(\bm{z} \!-\! \bm{h}(\bm{x}^{\star})\right) = \bm{\sigma}^{-2}\bm{H}^T\!\left[\bm{h}(\bm{x}_o) \!+\! \bm{\eta} \!+\! \bm{a} \!-\! \bm{h}(\bm{x}^{\star})\right],\nonumber\\
&\Rightarrow~\bm{H}^T\left[\bm{h}(\bm{x}_o) - \bm{h}(\bm{x}^{\star})\right] = -\bm{H}^T\left(\bm{\eta} + \bm{a}\right),\nonumber\\
% &\Rightarrow~\bm{H}^T\bm{H} \left(\bm{x}_o - \bm{x}^{\star}\right) \simeq -\bm{H}^T\left(\bm{\eta} + \bm{a}\right)\nonumber\\
&\Rightarrow~\left(\bm{x}_o - \bm{x}^{\star}\right) \simeq -\left(\bm{H}^T\bm{H}\right)^{-1}\bm{H}^T\left(\bm{\eta} + \bm{a}\right)
\label{eq:error}
\end{align}
where the last equation follows from \cref{eq:taylor}. Finally, from \cref{eq:wssr} and \cref{eq:error}, we have:
\begin{equation}
  \qz \simeq \sigma^{-2}\|\bm{P}(\bm{a} + \bm{\eta})\|^2,\label{eq:minWSSR}
\end{equation}
where $\bm{P} := \bm{I} - \bm{H}\left(\bm{H}^T\bm{H}\right)^{-1}\bm{H}^T = \bm{P}^2$ is the orthogonal projection (or hat) matrix. Note that this takes the same form as the residual in the linear case, albeit with the system matrix replaced by the Jacobian.

\section{Chi-square mechanism DP proof}\label{app:chi}
Letting $s := r + r'$, we have that:
\begin{equation}
\tqz \sim \chi^2_{s}(\theta^2) \quad\text{and}\quad \tilde{\mathbbm{q}}(\bm{z}') \sim \chi^2_{s}(\theta^2+\Delta_{\theta^2})
\end{equation}

The log-likelihood ratio of $\tqz$ and $\tilde{\mathbbm{q}}(\bm{z}')$ is given by:
\begin{align}
L & = \log\!\! \frac{f_{\tqz}(\tilde{q})}{f_{\tilde{\mathbbm{q}}(\bm{z}')}(\tilde{q})} \!=\! \log \!\frac{\frac{1}{2}e^{-(\tilde{q}+\theta^2)/2}\left(\frac{\tilde{q}}{\theta^2}\right)^{\frac{s}{4}-\frac{1}{2}}I_{\frac{s}{2}-1}(\sqrt{\theta^2\tilde{q}})}{\frac{1}{2}e^{-(\tilde{q}+{\theta'}^2)/2}\left(\frac{\tilde{q}}{{\theta'}^2}\right)^{\frac{s}{4}-\frac{1}{2}}I_{\frac{s}{2}-1}(\sqrt{{\theta'}^2\tilde{q}})}\nonumber \\
  & \!=\! \frac{{\theta'}^2\!-\!\theta^2}{2} + \left(\frac{s}{4}-\frac{1}{2}\right) \log \frac{{\theta'}^2}{\theta^2} + \log \frac{I_{\frac{s}{2}-1}(\sqrt{\theta^2\tilde{q}})}{I_{\frac{s}{2}-1}(\sqrt{{\theta'}^2\tilde{q}})},
\end{align}
where $I_a(b)$ is the modified Bessel function of the first kind.

At this stage, it is important to mention the following theorem on the ratio of modified Bessel functions of the first kind that was independently proved by authors of \cite{ross1972inequalities,bordelon1973inequalities}:
\begin{thm}
For all $a>0$ and $0<x<y$, the following inequalities hold:
\begin{equation}
  e^{x-y} \left(\frac{x}{y}\right)^{a} < \frac{I_a(x)}{I_a(y)} < e^{y-x} \left(\frac{x}{y}\right)^{a}.
\end{equation}
\end{thm}
Next, consider the event $|L| < \epsilon$. Its probability may be written as follows:
\begin{equation}
P\left(|L| \leq \epsilon\right) = P(L \leq \epsilon) - P(L \leq -\epsilon).\label{eq:Pdp}
\end{equation}
In order to find a lower bound for the probability of occurrence of this event, we shall find a lower bound for the first term and an upper bound for the second term in \cref{eq:Pdp}. We have for ${\theta'}^2 > \theta^2 > 0$:
\begin{align}
&P(L \leq \epsilon) = P\left[\left(\begin{aligned}
  &\frac{{\theta'}^2-\theta^2}{2} + \left(\frac{s}{4}-\frac{1}{2}\right) \log \frac{{\theta'}^2}{\theta^2} \\
  & \qquad + \log \frac{I_{\frac{s}{2}-1}(\sqrt{\theta^2\tilde{q}})}{I_{\frac{s}{2}-1}(\sqrt{{\theta'}^2\tilde{q}})} 
\end{aligned}\right)
\quad
\begin{aligned}
  \leq \epsilon
\end{aligned}\right]\nonumber                                                    \\
& \geq P\left[\left(\begin{aligned}
  &\frac{{\theta'}^2-\theta^2}{2} + \left(\frac{s}{4}-\frac{1}{2}\right) \log \frac{{\theta'}^2}{\theta^2} \\
  & + \log \left[e^{\sqrt{{\theta'}^2\tilde{q}} - \sqrt{\theta^2\tilde{q}}} \left(\frac{\sqrt{\theta^2\tilde{q}}}{\sqrt{{\theta'}^2\tilde{q}}}\right)^{\frac{s}{2}-1}\right] 
\end{aligned}\right)
\quad
\begin{aligned}
  \leq \epsilon
\end{aligned}\right]\nonumber                                                    \\
% &= P\left(\frac{{\theta'}^2-\theta^2}{2} + \left(\frac{s}{4}-\frac{1}{2}\right) \log \frac{{\theta'}^2}{\theta^2} \right.\nonumber\\
% &\qquad\qquad \left.+ \sqrt{{\theta'}^2\tilde{q}} - \sqrt{\theta^2\tilde{q}} + \left(\frac{s}{4}-\frac{1}{2}\right)\log\frac{\theta^2}{{\theta'}^2} \leq \epsilon\right)\nonumber\\
% &= P\left(\left(\theta' - \theta\right)\sqrt{\tilde{q}}  \leq \epsilon - \frac{{\theta'}^2-\theta^2}{2}\right)\nonumber\\
% &= P\left(\sqrt{\tilde{q}}  \leq \frac{\epsilon}{\theta' - \theta} - \frac{\theta' + \theta}{2}\right)\nonumber\\
 & = P\left(\tilde{q}  \leq \left(\epsilon/(\theta' - \theta) - (\theta' + \theta)/2\right)^2\right).
\end{align}
Similarly, an upper bound for the second term in \cref{eq:Pdp} is given by:
\begin{equation}
P(L \leq -\epsilon) \leq P\left(\tilde{q}  \geq \left(\epsilon/(\theta' - \theta) + (\theta' + \theta)/2\right)^2\right).
\end{equation}
Thus,
\begin{align}
P\left(|L| \leq \epsilon\right) & \geq P\left(\tilde{q}  \leq \left(\epsilon/(\theta' - \theta) - (\theta' + \theta)/2\right)^2\right) \nonumber   \\
                                & \qquad- P\left(\tilde{q}  \geq \left(\epsilon/(\theta' - \theta) + (\theta' + \theta)/2\right)^2\right)\nonumber \\
                                & =: 1 - \delta,
\end{align}

\section{Sensitivity Analysis}\label{app:sensitivity}
We are interested in the deviation in $\bm{P}$ when an element in $\bm{z}$ is changed. In order to find this deviation, we need to first write $\bm{P}'$ as a function of $\bm{P}$.
% \subsubsection{Rank Two correction of the System matrix}
% We have that $\bm{P}' = \bm{I} - \bm{H}'\bm{H}'^\dagger$, where $\bm{H}'^\dagger = \left(\bm{H}'^T\bm{H}'\right)^{-1}\bm{H}'^T$.
Since $\bm{H}^T\bm{e} = \bm{h}$ (the $m$\textsuperscript{th} row of $\bm{H}$), we can write the following:
\begin{align}
\bm{H}'&^T\bm{H}'  = \left(\bm{H}^T + \bm{\Delta}_h\bm{e}^T\right) \left(\bm{H} + \bm{e}\bm{\Delta}_h^T\right)\nonumber            \\
     & = \bm{C}_0 + \bm{H}^T\bm{e}\bm{\Delta}_h^T + \bm{\Delta}_h\bm{e}^T\bm{H} + \bm{\Delta}_h\bm{e}^T\bm{e}\bm{\Delta}_h^T\nonumber                                                    \\
     & = \bm{C}_0 + \bm{h}\bm{\Delta}_h^T + \bm{\Delta}_h\bm{h}^T + \bm{\Delta}_h\bm{\Delta}_h^T = \bm{C}_0 - \bm{h}\bm{h}^T + \bm{h}'\bm{h}'^T \nonumber                                                    \\
     &= \left(\bm{I} + \bm{C}_1\right)\bm{C}_0\left(\bm{I} + \bm{C}_1\right)^T = \bm{C}_2\bm{C}_0^{-1}\bm{C}_2^T,
\end{align}
where $\bm{C}_0 := \bm{H}^T\bm{H}$, $\bm{C}_1 := \bm{\Delta}_{h}\bm{h}^T\bm{C}_0^{-1}$ and $\bm{C}_2 := \left(\bm{C}_0 + \bm{\Delta}_{h}\bm{h}^T\right)$. 
Using the Sherman-Morrison formula, the inverse of $\bm{C}_2$ may be written as follows:
\begin{equation}
\bm{C}_2^{-1} = \left(\bm{C}_0 + \bm{\Delta}_{h}\bm{h}^T\right)^{-1} = \bm{C}_0^{-1}\left(\bm{I} - \frac{\bm{C}_1}{c_0}\right),
\end{equation}
where $c_0 := 1 + \bm{h}^T\bm{C}_0^{-1}\bm{\Delta}_{h}$. Consequently, the inverse of $\bm{H}'^T\bm{H}'$ is:
\begin{equation}
  \left(\bm{H}'^T\bm{H}'\right)^{-1} = \bm{C}_2^{-T}\bm{C}_0\bm{C}_2^{-1} 
  % &= \left(\bm{I} - \frac{\bm{C}_1^T}{c_0}\right)\bm{C}_0^{-1}\left(\bm{I} - \frac{\bm{C}_1}{c_0}\right)\nonumber\\
 % & = \left[\bm{C}_0^{-1} - \frac{\bm{C}_1^T\bm{C}_0^{-1}}{c_0}\right]\left(\bm{I} - \frac{\bm{C}_1}{c_0}\right)\nonumber\\
 % & = \bm{C}_0^{-1} + \left[- \frac{\bm{C}_0^{-1}\bm{C}_1}{c_0} - \frac{\bm{C}_1^T\bm{C}_0^{-1}}{c_0}\right.\nonumber\\
 % & \qquad \qquad\qquad\qquad+  \left.\frac{\bm{C}_1^T\bm{C}_0^{-1}\bm{C}_1}{c_0^2}\right]\label{eq:C1} \\
 = \bm{C}_0^{-1} + \bm{C}_3,
\end{equation}
where 
\begin{equation}
  \bm{C}_3 := -\frac{\bm{C}_0^{-1}\bm{C}_1}{c_0} - \frac{\bm{C}_1^T\bm{C}_0^{-1}}{c_0} +  \frac{\bm{C}_1^T\bm{C}_0^{-1}\bm{C}_1}{c_0^2}
\end{equation}
Finally, we can write $\bm{P}'$ in terms of $\bm{P}$:
\begin{align}
\bm{P}' & = \bm{I} - \bm{H}'\bm{H}'^\dagger = \bm{I} - \bm{H}'\left(\bm{H}'^T\bm{H}'\right)^{-1}\bm{H}'^T \nonumber                                                                  \\
        & = \bm{I} - \left(\bm{H} + \bm{e}\bm{\Delta}_{h}^T\right) \left(\bm{C}_0^{-1} + \bm{C}_3\right) \left(\bm{H}^T + \bm{\Delta}_{h}\bm{e}^T\right)\nonumber \\
        & = \bm{I} - \bm{H}\bm{C}_0^{-1}\bm{H}^T + \bm{C}_4 = \bm{P} + \bm{C}_4,
\end{align}
where
\begin{align}
 & \bm{C}_4 := - \bm{H}\bm{C}_0^{-1}\bm{\Delta}_{h}\bm{e}^T\nonumber                                                                              \\
 & - \left[\bm{H}\bm{C}_3 + \bm{e}\bm{\Delta}_{h}^T\left(\bm{C}_0^{-1} +\bm{C}_3\right) \right] \left(\bm{H}^T + \bm{\Delta}_{h}\bm{e}^T\right)
\end{align}
is the correction in $\bm{P}$ if $\bm{H}$ is rank two corrected.

\section{Normal Approximation Proof}
Let $\bm{H} = \bm{U}\bm{\Sigma}\bm{V}^T$ be the SVD of $\bm{H}$, where $\bm{U} = [\bm{u}_1, \ldots,\bm{u}_m]\in\mathbb{R}^{m \times m}, \bm{V} = [\bm{v}_1,\ldots,\bm{v}_n]\in\mathbb{R}^{n \times n}$. Then, the RWLS model's state estimate in \cref{eq:reg_state} can be rewritten as:
\begin{equation}
\bm{x}^{\star} = \bm{V}\left(\lambda\sigma^2\bm{I}+\bm{\Sigma}^T\bm{\Sigma}\right)^{-1}\bm{\Sigma}^T\bm{U}^T\bm{z}
\end{equation}
and $\bm{P}_{\lambda}$ can be rewritten as:
\begin{equation}
\bm{P}_{\lambda} = \bm{U}\left(\bm{I} - \bm{\Sigma}\left(\bm{\Sigma}^T\bm{\Sigma} + \lambda\sigma^2 \bm{I}\right)^{-1}\bm{\Sigma}^T\right)\bm{U}^T.
\end{equation}
Thus, the WSSR is given by:
\begin{equation}
\qz = \frac{\bm{z}^T\bm{U}\bm{D}\bm{U}^T\bm{z}}{\sigma^2} = \bm{z}_U^T\bm{D}\bm{z}_U = \sum_{i=1}^m  D_{ii} z_{U,i}^2,
\end{equation}
where $\bm{D} := \left(\bm{I} - \bm{\Sigma}\left(\lambda\sigma^2\bm{I}+\bm{\Sigma}^T\bm{\Sigma}\right)^{-1}\bm{\Sigma}^T\right)^2$, $\bm{z}_U := \frac{\bm{U}^T\bm{z}}{\sigma} \sim \mathcal{N}(\bm{\theta}, \bm{I})$, and $\bm{\theta} := \bm{\Sigma}\bm{V}^T\bm{x}+\bm{U}^T\bm{a}$. This implies that:
\begin{equation}
z_{U,i}^2 \stackrel{ind}{\sim} \chi_1^2(\theta_i^2), \qquad \forall i \in [m].
\end{equation}
Thus, the WSSR is a random variable of chi-squared-type mixtures. The theorem then follows from \cite[Theorem 1]{zhang2005approximate}.

\end{appendices}
\end{document}